\documentclass[]{aa} 
\usepackage{graphicx}

\begin{document}
   \title{FUSE observations of H$_2$ around the Herbig AeBe stars \\ HD\,100546 and HD\,163296}
   \titlerunning{FUSE observations of H$_2$ around HD\,100546 and HD\,163296}

   \author{
A.~Lecavelier des Etangs \inst{1}
          \and
M.~Deleuil \inst{2}
          \and
A.~Vidal-Madjar \inst{1}
          \and
A.~Roberge \inst{3, 4}
          \and
F.~Le~Petit \inst{5}
          \and
G.~H\'ebrard \inst{1}
          \and
R.~Ferlet \inst{1}
          \and
P.\,D.~Feldman \inst{3}
          \and
J.-M.~D\'esert \inst{1}
          \and
J.-C.~Bouret \inst{2}
          }

   \offprints{A. Lecavelier des Etangs,
   \email{lecaveli@iap.fr}}

   \institute{Institut d'Astrophysique de Paris, CNRS, 98 bis Bld Arago, 
              Paris, France
         \and
               Laboratoire d'Astrophysique de Marseille, Marseille, France
         \and
               Department of Physics and Astronomy, Johns Hopkins University, Baltimore, MD, USA
         \and
               Current Address: Carnegie Institution of Washington, Washington, DC, USA
         \and
               Meudon Observatory, Meudon, France
  }

   \date{Received March 25, 2003; accepted June 12, 2003}

   \abstract{
We present the analysis of {\sl FUSE} observations of two Herbig AeBe stars 
known to harbor young circumstellar disks: HD\,100546 and HD\,163296.
In both cases we detect absorption lines from warm and dense H$_2$.
The thermalization of the rotational levels up to $J\sim 4$ 
allows evaluation of the temperature, density and typical size 
of the absorbing layer.
These quantities are consistent with absorption of the light 
of the central star by a thin layer of a circumstellar disk
seen at an intermediate inclination. 
   }

   \maketitle

\section{Introduction}
\label{introduction}

HD\,100546 and HD\,163296 are two Herbig AeBe stars harboring young circumstellar disks. 
Herbig AeBe stars are intermediate-mass (2--10~M$_{\sun}$) young 
emission-line stars still in the pre-main sequence phase (Herbig~\cite{Herbig1960}).
Material surrounding Herbig stars is thus considered as the constituent of
protoplanetary disks,
similar to the disks seen around the less massive T-Tauri stars. These disks are
the material from which (and the places where) planets are supposed to form according 
to the standard model of planetary formation (Lissauer~\cite{Lissauer1993}).

HD\,100546 (B9Vne, $d$=103$^{+7}_{-6}$\,pc, $t$$>$10\,Myr; 
van den Ancker et al.~\cite{Ancker1998}) 
is older than HD\,163296 
(A1Ve, $d$=122$^{+17}_{-13}$\,pc, $t$=$4^{+6}_{-2.5}$\,Myr; 
van den Ancker et al.~\cite{Ancker1998}). 
However, both stars present very similar
characteristics. In the two cases, ISO detected silicate features resembling cometary 
materials (Malfait et al.~\cite{Malfait1998}; van den Ancker et al.~\cite{Ancker2000}; 
Bouwman et al.~\cite{Bouwman2000}, \cite{Bouwman2001}).
Disks have recently been imaged around both stars (Pantin et al.~\cite{pantin};
Grady et al.~\cite{grady2000}, \cite{grady2001}; Augereau et al.~\cite{augereau}). 
The images show dusty disks at intermediate inclination ($\sim$60\degr).
Absorption spectroscopy also revealed the presence of circumstellar gas around HD\,100546,
including spectral variabilities which are interpreted as accreting material 
(Grady et al.~\cite{grady1996}, \cite{Grady1997}).

In the case of HD\,163296, a CO disk of about 310~AU in semi-major axis 
has been detected (Mannings \& Sargent~\cite{Mannings}).
H$_2$ infrared emission lines have also been reported by Thi et al.~(\cite{thi})
at 17 and 28$\mu$m.
From their ISO observations and assuming thermal equilibrium up to $J=3$, 
they derive a total H$_2$ mass
of $0.4(\pm 0.2)\times 10^{-3}$\,M$_{\sun}$, 
and the ratio of the two emission lines 
($v=0$, $J$=2$\rightarrow$0 and $J$=3$\rightarrow$1) gives a temperature of 
$220(\pm 22)$\,K. However, this detection has been challenged by more recent ground based 
observation at 17$\mu$m which shows no detection with three times better sensitivity 
(Richter et al.~\cite{richter}). 

These controversial observations raise the question of the molecular content of such
disks, where most of the mass is in the form of molecules, and particularly H$_2$.
The formation of giant planets requires a large reservoir of molecular gas. 
Moreover some of these planets migrate close to their parents where they are observed.
This migration also needs a massive disk to allow the angular momentum exchange
between the migrating planet and the disk.
The H$_2$ content is thus a key ingredient in the recipe for the formation of
giant planets. But H$_2$ is a symetrical molecule, and infrared emission 
by quadrupole rotational transitions is very inefficient. In contrast, 
when seen against a UV bright source, far-UV absorption lines due to 
electronic transitions allow sensitive 
observation of H$_2$ even at low temperatures. 
{\sl FUSE} offers a unique opportunity to scrutinize 
in detail the H$_2$ content of protoplanetary disks around Herbig stars.

\section{Observation and data analysis}

HD\,100546 was observed with {\sl FUSE} through the LWRS aperture ($30\arcsec$$\times$$30\arcsec$)
for a total time of 5.8~hours on March 26, 2000 (Program P1190303) 
and March 3, 2002 (Program P2190401) 
(for an overview of {\sl FUSE}, see Moos et al.~\cite{moos} and
Sahnow et al.~\cite{sahnow}).
HD\,163296 was observed twice, on April 27, 2001 and April 29, 2001 for
a total time of 8.9~hours (Programs P2190601 and Q2190101, respectively). 
The data of both targets were reprocessed with the version 2.0.5 of the 
{\tt CALFUSE}
pipeline. The output of the pipeline is a total of 5 and 10~sub-exposures 
for HD\,100546 
and HD\,163296, respectively. The sub-exposures have been aligned and 
coadded resulting
in a set of four independent spectra, one for each {\sl FUSE} channel 
(2 LiF spectra and 2 SiC spectra).

The version 2.0.5 of the {\tt CALFUSE} pipeline is known to slightly 
over-estimate the tabulated errors on each pixel. We compared the data 
used in the present work with data of the same observations but obtained 
with the version 2.2.1 of the {\tt CALFUSE} pipeline; we conclude
that the error bars given below are not significantly affected by the 
improvement in the error propagation of the different pipelines.

Apart from the observation of the molecular hydrogen, the spectra are 
rich in emission and absorption lines from atomic and ionic species.
For instance, a large number of Fe\,{\sc ii} lines from the ground level as well
as from excited levels are clearly detected. 
As already observed in AB~Aur and $\beta$~Pic (Roberge et al.~\cite{Roberge2001};
Deleuil et al.~\cite{deleuil2001}), bright emission from
C\,{\sc iii} and O\,{\sc vi} are also detected in the two spectra. 
All these features indicate the presence of circumstellar material, chromospheric activity
and/or accretion (Bouret et al.~\cite{bouret2002}).
A detailed analysis of the whole {\sl FUSE} spectrum will be made
in a forthcoming paper (Deleuil et al.~\cite{deleuil2003}).
Here we focus on the analysis of the H$_2$ lines, 
probing the molecular portion of the disk.

   \begin{figure}[tbp]
   \centering
   \includegraphics[height=\columnwidth,angle=-90]{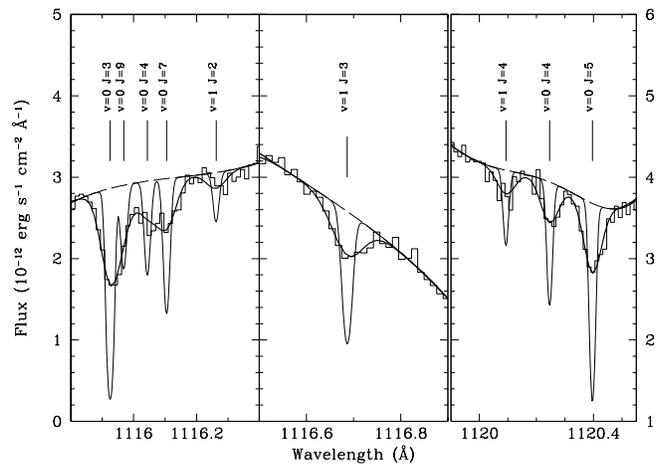}
      \caption{
Sample of H$_2$ lines detected toward HD\,100546. The histogram shows the 
data. The final fit to the data is shown (thick line) together with
the theoretical spectrum without convolution with the instrumental 
line spread function (thin line). This shows that the fitted lines
are not saturated. In the selected sample, some lines are blended 
together. Consequently, a profile fitting is needed to use the information 
included in the data but which can not be translated into 
an equivalent width.}
    \label{fit1}
    \end{figure}

   \begin{figure}[htbp]
   \centering
   \includegraphics[height=\columnwidth]{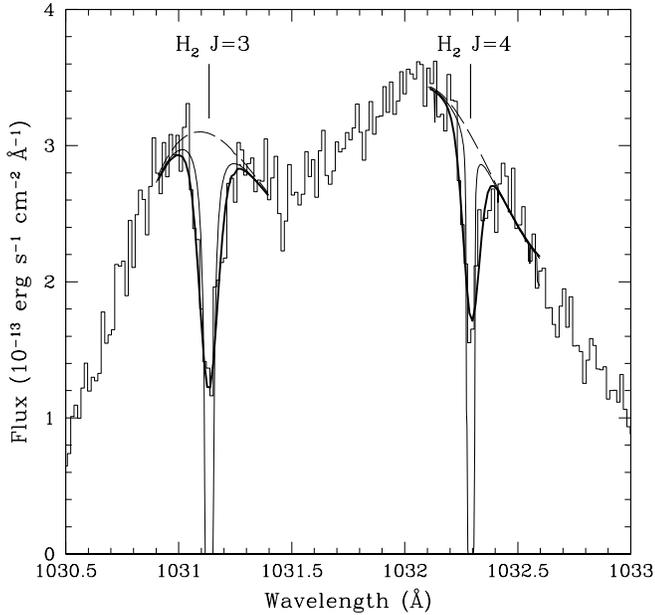}
      \caption{
Two H$_2$ absorption lines detected over one emission line of the 
O\,{\sc vi} doublet toward HD\,163296.
Here the theoretical spectrum without convolution with the instrumental 
line spread function (thin line as in Fig.~1) shows that these H$_2$ 
absorption lines are saturated.}
    \label{fit2}
    \end{figure}

In the far-UV, HD\,100546 is brighter than HD\,163296 
and the observed spectrum has a better S/N ratio allowing detection of 
H$_2$ in its pure rotational level ($v=0$) up to $J=9$
(a line of $v=0$, $J=10$ is also marginally detected at 1058.6~\AA) and
in its first vibrational level ($v=1$) up to $J=5$ (Fig.~\ref{fit1}).
In the HD\,163296 spectrum, H$_2$ lines are detected up to $v=0$, $J=4$. 
Higher $J$-levels are beyond the detection limit.
The data analysis and profile fitting has been done using the 
{\tt Owens} code kindly made available to us by Dr.~M.~Lemoine 
(see for example Lemoine et al.~\cite{lemoine} and H\'ebrard et al.~\cite{hebrard}).
For the electronic transitions, we used the wavelengths and oscillator strengths 
tabulated by Abgrall et al. (\cite{abgrall1993a} and \cite{abgrall1993b} for the Lyman 
and the Werner system, respectively), and the inverses of the total radiative lifetimes 
tabulated by Abgrall et al. (\cite{abgrall2000}).

   \begin{figure}[htbp]
   \centering
   \includegraphics[width=\columnwidth]{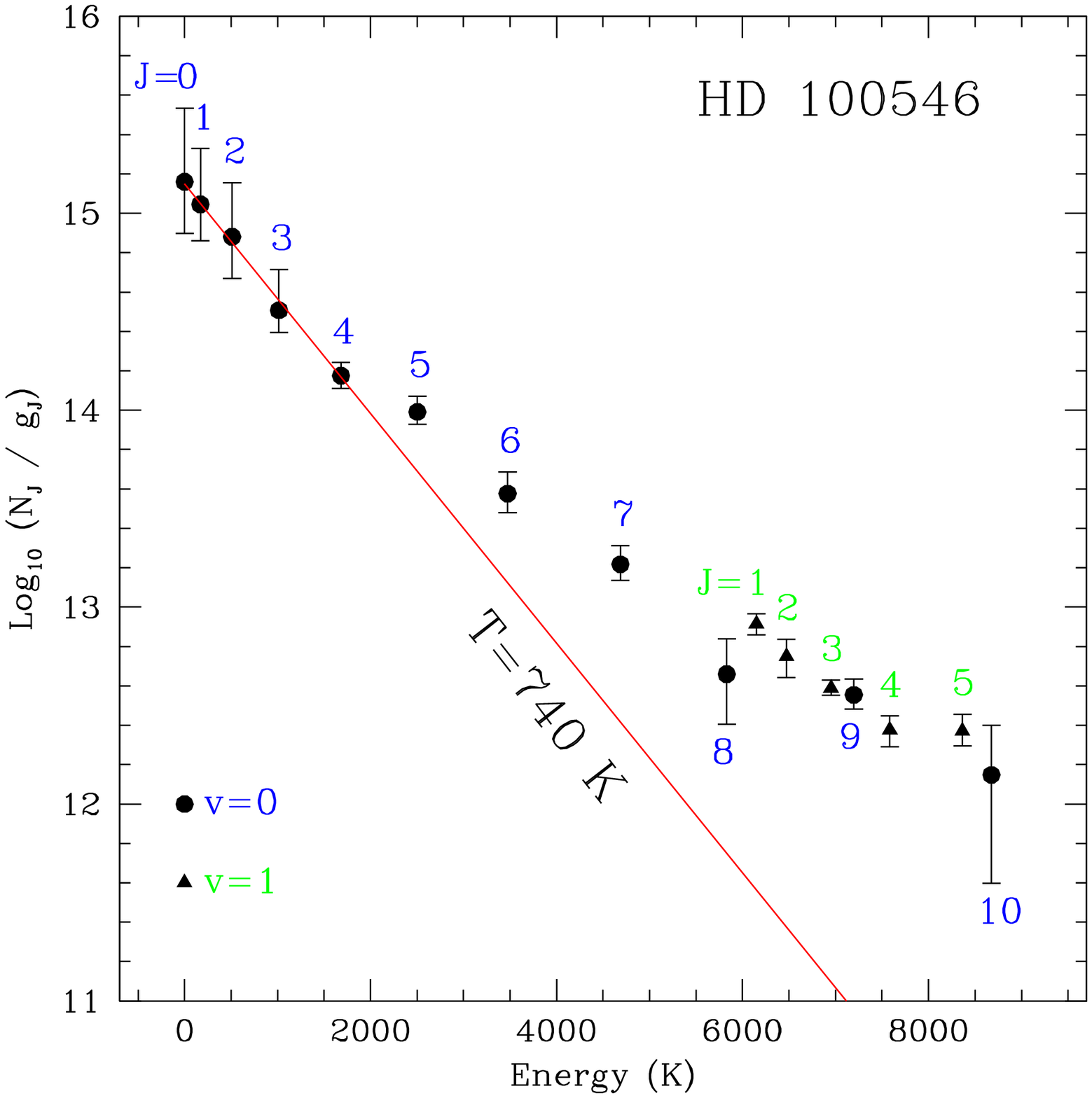}
      \caption{
Excitation diagram of the H$_2$\ lines detected toward HD\,100546. 
The column densities of the rotational levels with $J\le 4$ are consistent 
with a single temperature of $T\sim 740$~K. This shows that the
H$_2$\ volume density must be higher than the critical value of
$n_{\rm H_2}\approx 10^4$cm$^{-3}$.
              }
    \label{excitation plot 100546}
    \end{figure}

   \begin{figure}[htbp]
   \centering
   \includegraphics[width=\columnwidth]{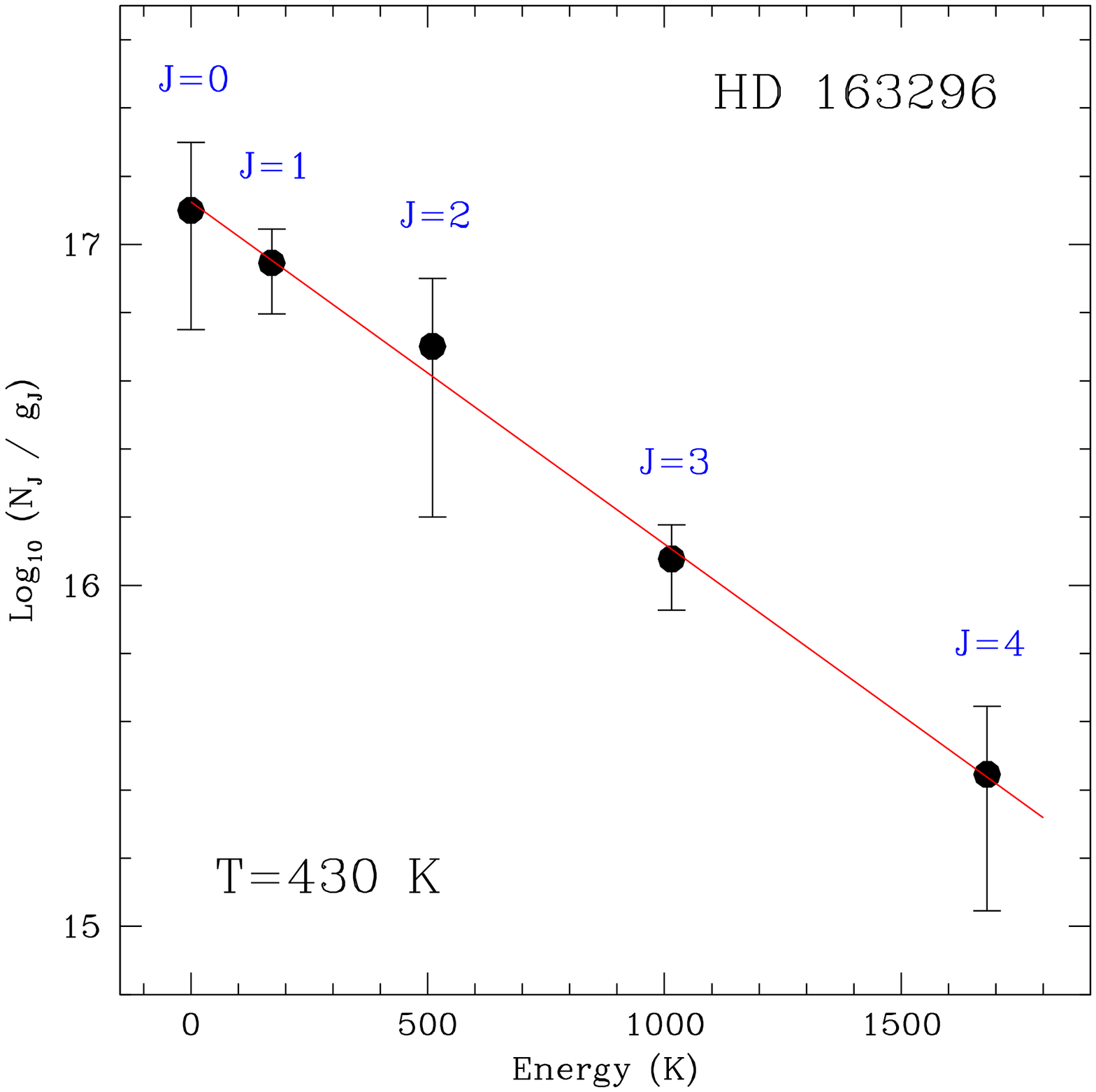}
      \caption{
Excitation diagram of the H$_2$\ lines detected toward HD\,163296. 
The column densities of all the detected $J$-levels are consistent 
with a single temperature of $T\sim 430$~K. This shows that the
H$_2$\ volume density must be higher than the critical value of
$n_{\rm H_2}\approx 8\times 10^4$cm$^{-3}$.
              }
    \label{excitation plot 163296}
    \end{figure}

With low H$_2$ column densities
in HD\,100546 ($N\le 10^{16}$cm$^{-2}$ for each $J$-level), we can 
choose to fit only the unsaturated lines having low oscilator strength
($f \la 10^{-2}$). This allows us to avoid systematic
errors in the estimates of column densities which could be included by 
the fit of saturated lines with uncertain instrumental line 
spread function. Note, however, that additional fits have been performed
with saturated lines ($f\ga 10^{-2}$).
They give very similar column densities
and allow the determination of the intrinsic line width ($b$). 
In HD\,163296
the column densities are larger, and all the H$_2$ lines are saturated. 
Therefore the error bars on the column densities are also larger than
in HD\,100546.
An example of the fit of two H$_2$ lines ($J=3$ and $J=4$) which 
are superimposed
on the O\,{\sc vi} stellar emission line is given in Fig.~\ref{fit2}.

   \begin{table}[htbp]
      \caption[]{H$_2$ column density toward HD\,100546 and HD\,163296}
         \label{N_H2}
         \begin{tabular}{lllllll}

            \hline
            \noalign{\smallskip}
            \multicolumn{3}{c}{HD\,100546} & \hspace{2cm} &  \multicolumn{3}{c}{HD\,163296} \\
            \noalign{\smallskip}
            \hline
            \noalign{\smallskip}
            $J$  & $v$      &  log $N_{\rm H_2}$ $^{\rm a}$ & \hspace{2cm} & 
            $J$  & $v$      &  log $N_{\rm H_2}$ $^{\rm a}$ \\
                 &          &  (cm$^{-2}$)                  & \hspace{2cm} &
                 &          &  (cm$^{-2}$)                  \\
            \noalign{\smallskip}
            \hline
            \noalign{\smallskip}
0    & 0 &  15.16$^{+0.38}_{-0.26}$    & \hspace{2cm} & 0    & 0 &  17.1$^{+0.4}_{-0.7}$    \\
1    & 0 &  16.00$^{+0.28}_{-0.19}$    & \hspace{2cm} & 1    & 0 &  17.9$^{+0.2}_{-0.3}$    \\
2    & 0 &  15.58$^{+0.28}_{-0.21}$    & \hspace{2cm} & 2    & 0 &  17.4$^{+0.4}_{-1.0}$    \\
3    & 0 &  15.83$^{+0.21}_{-0.11}$    & \hspace{2cm} & 3    & 0 &  17.4$^{+0.2}_{-0.3}$    \\
4    & 0 &  15.13$^{+0.07}_{-0.07}$    & \hspace{2cm} & 4    & 0 &  16.4$^{+0.4}_{-0.8}$    \\
5    & 0 &  15.51$^{+0.08}_{-0.06}$    \\
6    & 0 &  14.69$^{+0.11}_{-0.10}$    \\
7    & 0 &  14.87$^{+0.10}_{-0.08}$    \\
8    & 0 &  13.89$^{+0.18}_{-0.25}$    \\
9    & 0 &  14.31$^{+0.08}_{-0.07}$    \\
10   & 0 &  13.47$^{+0.25}_{-0.55}$    & (3$\sigma$ detection) \\
            \noalign{\smallskip}
            \noalign{\smallskip}
1    & 1 &  13.87$^{+0.05}_{-0.06}$    \\
2    & 1 &  13.45$^{+0.09}_{-0.11}$    \\
3    & 1 &  13.91$^{+0.04}_{-0.04}$    \\
4    & 1 &  13.33$^{+0.07}_{-0.09}$    \\
5    & 1 &  13.89$^{+0.08}_{-0.08}$    \\
            \noalign{\smallskip}
            \hline
         \end{tabular}
\begin{list}{}{}
\item[$^{\mathrm{a}}$] The error bars are 2$-\sigma$ error bars.
\end{list}
   \end{table}

The estimated column densities are tabulated with $2\sigma$ error bars 
(Table~\ref{N_H2}). 
We obtain the total
column densities 
$$
N_{\rm H_2}({\rm HD\,100546}) \approx  2.8\times 10^{16}{\rm cm}^{-2}
$$
and 
$$
N_{\rm H_2}({\rm HD\,163296}) \approx  1.4\times 10^{18}{\rm cm}^{-2}.
$$

The error bars are estimated by the classical method of the $\Delta \chi^2$ 
increase of the $\chi^2$ of the fit
(see H\'ebrard et al.~(\cite{hebrard}) for a full discussion on the fit 
method and error estimates with {\tt owens} profile fitting). 
These error bars include
the uncertainties in the continuum fits, intrinsic line widths ($b$)
and the instrumental line spread function.
We note that the observation of different lines with different 
oscillator strengths for the same species allows us to constrain
all these quantities. In particular the fit to saturated lines 
constrains the line widths and the instrumental line spread function.
This last one is allowed to slowly vary with the wavelength. 
The results are
obtained from a final self-consistent fit to all the data in which the
continuum, the instrumental line spread function, and physical
parameters of the absorption lines are free parameters 
and estimated by the
determination of the best $\chi^2$ in the parameters' space.
For the intrinsic line widths, we obtain 
$b_{\rm HD\,100546}=3.4^{+0.8}_{-0.5}$km~s$^{-1}$ and 
$b_{\rm HD\,163296}=2.2^{+0.5}_{-0.7}$km~s$^{-1}$. 

\section{Results}

\subsection{Excitation diagrams}

The excitation diagrams corresponding to the derived H$_2$ column densities
are presented in Fig.~\ref{excitation plot 100546} and~\ref{excitation plot 163296},
for HD\,100546 and HD\,163296, respectively.
For the $v=1$ levels observed toward HD\,100546, we see that the column densities are above 
a simple extrapolation 
of the $v=0$ levels as plotted on the excitation plot (Fig.~\ref{excitation plot 100546}). 
This effect is also observed in the vibrationally excited interstellar H$_2$ detected 
toward HD~37903 (Meyer et al.~\cite{meyer}),
where the high energy levels are not thermally populated.
Note also that the level degeneracies ($g$)
of the $v=1$ levels are smaller than the degeneracies of 
the $v=0$ levels with similar energy, hence in the same region of the plot.
The plotted values of $N/g$ are thus larger for these $v=1$ levels 
than for the nearby $v=0$ levels (Fig.~\ref{excitation plot 100546}).

\subsection{Turbulence}

The simultaneous determination of temperature and line broadening can 
be used to constrain the turbulence of the gas.
The observed line width $b$ is a combination of the thermal broadening, 
the intrinsic turbulence $v_{\rm turb}$, and the projected radial component 
of the gas motion seen along the star diameter.
With the knowledge of the temperature, and considering that 
the motion can only increase the line width, we can estimate a lower
limit on the turbulence velocity of the gas.
We have
$$
b \ge \sqrt{
 \frac{2kT}{\mu_{\rm H_2}} + 
 v_{\rm turb}^2      }
$$
where $\mu_{\rm H_2}$ is mass of molecular hydrogen.
This can be written as
$$
 \left(\frac{v_{\rm turb}}{ {\rm km\,s}^{-1}}\right)^2 
\le 
\left(\frac{b}{ {\rm km\,s}^{-1}}\right)^2 
-
8.31\times 10^{-3}\frac{T}{{\rm K}}.
$$
Using the temperature given above and the 2$\sigma$ upper limits
on the measured line widths ($b^{2\sigma}_{\rm HD\,100546}\le 4.2$\,km\,s$^{-1}$ and 
$b^{2\sigma}_{\rm HD\,163296}\le 2.7$\,km\,s$^{-1}$), we find that
$v_{\rm turb, HD\,100546}$$\le$$3.4$\,km\,s$^{-1}$ and 
$v_{\rm turb, HD\,163296}$$\le$$1.9$\,km\,s$^{-1}$.

\subsection{Nature and origin of the H$_2$ gas}

Many characteristics show that the detected H$_2$ is not interstellar but 
circumstellar; this favours the interpretation that
the detected lines arise from protoplanetary material. 
First, in both cases the radial velocity of H$_2$ is similar to the 
radial velocity of the atomic lines due to the circumstellar gas.
Lines from exited levels of Fe\,{\sc ii} 
(like Fe\,{\sc ii}* and Fe\,{\sc ii}**, Deleuil et al.~\cite{deleuil2003})
cannot be due to the interstellar medium. They must be linked to
dense circumstellar material.
The similarity of their
radial velocity with the radial velocity of H$_2$ is a first clue that
the H$_2$ is linked with the circumstellar matter.

Importantly, the excitation diagrams show 
that the H$_2$ is thermalized up to about $J=4$, with high temperatures.
Hot H$_2$ has also been observed around the T-Tauri star
TW~Hya ($T$=3000~K, Herczeg et al.~\cite{herczeg2002}).
Here we obtain
$$T\approx 740\pm 30 ~{\rm K}\ \  {\rm for~HD\,100546}$$ 
$$ {\rm and} \ \ \ T\approx 430\pm 20 ~{\rm K}\ \  {\rm for~HD\,163296}.$$
Seen in absorption,
it is very unusual to observe such a thermalization at high temperature up to $J=4$ 
(to our knowledge, this is the first case).
This shows that the absorbing medium has very particular physical conditions, 
different from what is commonly seen in the diffuse interstellar medium.
Indeed this requires a line of sight which presents simultaneously 
high density for the thermalization and low thickness to have column densities 
low enough for a detection of the far-UV stellar continuum.
In the diffuse and transluscent interstellar media, levels higher than $J=1$
or $J=2$ are populated by mechanisms other than the collisional pumping.
In these low density media, the collisional pumping is less efficient 
than the UV-pumping and mecanical processes such as C~shocs or turbulence. 
Here levels up to $J=4$ show a dominant collisional pumping.
This is a clue of a high density medium.

Note here that for the levels higher than $J=4$ as detected in HD\,100546, 
it is not possible to discriminate between a collisional 
pumping within a higher temperature component of the absorber and other
mecanical or radiative processes.

The observed high temperatures give additional clues that the detected H$_2$ 
is circumstellar and close to the exciting stars. Using the numerical model
developed by Le Bourlot et al.~(\cite{lebourlot1993}), 
we find that the incident radiation on the H$_2$ must be $\sim 10^4$ times larger 
than the mean galactic UV radiation field.
This shows that the observed H$_2$ is circumstellar and really close to its
central star.
These temperatures 
correspond to $\sim$1.5~AU from HD\,100546 and $\sim$4~AU from HD\,163296.

Most importantly, the thermalization allows us to estimate the H$_2$ volume density 
($n_{\rm H_2}$). 
Using the critical density given by 
Le Bourlot et al.~(\cite{lebourlot1999}), we find :
$$n_{\rm H_2}({\rm HD\,100546}) \ga  10^4{\rm cm}^{-3}$$
$$ {\rm and} \ \ \ n_{\rm H_2}({\rm HD\,163296}) \ga  8\times 10^4{\rm cm}^{-3}.$$

Assuming these volume densities, we can estimate the typical size of the absorbing layer 
responsible for the observed absorptions.
Note that with low column densities and large volume densities,
the absorbing cloud must be thin. The ratios of the H$_2$ column
densities to the volume densities give a typical thickness 
of $\la$0.2~AU for HD\,100546 and $\la$1.2~AU for HD\,163296. \\

All the observed and estimated quantities are consistent with absorption 
of the star light by a thin outer layer of a disk seen 
at an intermediate inclination.
Because the disks of HD\,100546 and HD\,163296 are both inclined by about 60\degr,
it is likely that we observed a line of sight grazing the disk surface. In any
case, this shows that the H$_2$ gas is not confined to a very flat disk.
Alternatively, there is still a possibility that there might be shocked
molecular gas in the vicinity of these stars, due to the combination 
of a bipolar jet (as observed around HD\,163296) and a circumstellar envelope.
However it is not possible to discriminate between these two possibilities
with solely spectroscopic observation with no spatial information.

\section{Conclusions}

We observed the molecular hydrogen toward two Herbig stars
surrounded by young protoplanetary disks. 
The observation of an extremely large number of rotational and vibrational
levels allows for the first time the determination of the H$_2$ physical 
conditions within circumstellar gas around two HAeBe stars.
With this detailed view of the population of the H$_2$ excited levels,
we now need a full modeling in which all the cooling and heating terms must
be taken into account. Such a modeling is under development.
This opens a new window into the physical conditions operating in the 
protoplanetary disks surrounding massive stars.

The present observations of material surrounding stars with typical age 
between 10$^6$ and 10$^7$~years
give new constrains on the relation between the radiation and 
the molecular protoplanetary 
disk in which planets are supposed to form and migrate 
(Terquem et al.~\cite{Terquem2000}).
This gives information on the interface between the dense part of the disks
and the interstellar medium, and can help to better understand the late stages
of the planetary formation. The radiation is indeed a key element
in the evaporation of the protoplanetary disks whose lifetime seems to be
limited (Zuckerman et al.~\cite{Zuckerman1995}) 
although the process responsible for the clearing of the disks remains unclear.

\begin{acknowledgements}
This work has been done using the profile fitting procedure 
developed by M.~Lemoine and the {\sl FUSE} French Team. 
The data were obtained for the Guaranteed Time Team by the
NASA-CNES-CSA {\sl FUSE} mission operated by the Johns Hopkins University.
Financial support to French participants has been provided by CNES.
Financial support to U.\,S. participants has been provided by NASA
contract NAS5-32985.
We warmly thank E. Roueff 
for providing H$_2$ transition data in electronic format and for fruitful
discussions. We thank A.~Dutrey, G.~Herczeg and J.~Linsky for 
fruitful comments.

\end{acknowledgements}

\end{document}